\documentclass[aps,pre,reprint, amsmath, amssymb,superscriptaddress]{revtex4-1}

\usepackage{morefloats}
\usepackage{bm}
\newcommand{\beq}{\begin{equation}}
\newcommand{\eeq}{\end{equation}}
\newcommand{\kB}{k_{\mbox{\tiny B}}}
\usepackage[retainorgcmds]{IEEEtrantools}
\usepackage{graphicx,tikz,placeins}
\usepackage{mathrsfs}
\usepackage{amsmath,amssymb,amsfonts,physics}
\usepackage{color}
\usepackage{float}
\usepackage{times,txfonts}
\usepackage{nicefrac}
\usepackage[colorlinks=true,linkcolor=blue,urlcolor=blue,citecolor=blue,pdfusetitle]{hyperref}
\usepackage{physics}
\usepackage{soul}
\newcommand{\mom}[1]{\left\langle#1\right\rangle}

\begin{document}

\title{ Obtaining efficient  collisional engines via velocity dependent drivings}

\author{Iago N. Mamede, Angel L. L. Stable and C. E. Fiore}
\affiliation{Universidade de S\~ao Paulo,
Instituto de F\'isica,
Rua do Mat\~ao, 1371, 05508-090
S\~ao Paulo, SP, Brazil}
\date{\today}
\begin{abstract}
Brownian particles interacting sequentially with  distinct  temperatures
and  driving forces at each stroke have been tackled as  a reliable alternative for the construction of  engine setups.   However they can behave very inefficiently depending on  the driving used for the worksource  and/or when temperatures of each stage are very different from each other. Inspired by some models for molecular motors
and recent experimental studies, a coupling
between driving and velocities is introduced as an alternative ingredient for enhancing the system performance.
Here, the role of this new  ingredient for levering the engine performance is detailed  investigated from  stochastic thermodynamics.
Exact expressions for quantities and distinct maximization routes have been obtained and investigated.
The search of an optimal coupling   provides  a substantial increase of engine performance (mainly efficiency),  even for large $\Delta T$.
A  simple and general argument for the optimal coupling can be estimated, irrespective the driving and other model details.
\end{abstract}

\maketitle

\section{Introduction}
One of the main  goals of nonequilibrium thermodynamics is to understand, from an operational point of view, the  conversion  between distinct amounts of energy  delivered  and those converted into useful power output \cite{callen1960thermodynamics}. Such fundamental issue appears in several
systems in nature, encompassing physical \cite{PhysRevLett.95.190602,seifert2012stochastic,rana2014single,martinez2016brownian}, biological \cite{liepelt1,liepelt2}, chemical processes \cite{seader1982thermodynamic}, quantum technologies and others, 
 thereby illustrating the great deal of attention  for describing thermal machines  operating at the nanoscale \cite{PhysRevLett.95.190602,seifert2012stochastic}. Among
the  setups, we cite those   composed of quantum-dots \cite{PhysRevE.81.041106}, colloidal particles \cite{rana2014single,martinez2016brownian, albay2021shift}, single \cite{proesmans2016brownian} and coupled systems \cite{PhysRevLett.109.190602,mamede2021obtaining} acting as working substance and others \cite{jun2014high}. 
Most of above examples deal with 
engines operating under fixed 
or  time-periodic variation of external parameters.

Collisional  machines has also been tackled as a candidate for reliable thermal engines, in which
the system is sequentially exposed to  a distinct thermal reservoir and external driving
forces and the time required for switching the  thermal baths at the end of each stage being neglected.
Despite its reliability in distinct situations, encompassing
systems  interacting  only with a small fraction of the environment and those
presenting distinct drivings over each member of system \cite{benn1,maru,saga,parrondo},
such class of systems can operate  inefficiently depending on the way it is projected (temperatures, kind of driving and duration of each stroke). For this reason, recent strategies, such an optimal switching time between thermal baths \cite{PhysRevResearch.3.023194,noa2021efficient} 
and the choice of an appropriate driving \cite{akasaki2022thermodynamics} at each stroke have been proposed and investigated.  
However  about improvements can be limited
when  heat can not be converted into output work and 
the temperature difference $\Delta T$ between strokes increases, yielding
 small efficiencies \cite{noa2020thermodynamics,noa2021efficient,akasaki2022thermodynamics}. 


Aimed at  circumventing above limitation,  we introduce  a new  ingredient  as an strategy for improving
the efficiency of thermal engines.  It consists of including a velocity dependent  driving 
resulting in the generation of output power due to
two component drivings: the first, given by $f_ih_i(t)$,  coming from
an arbitrary driving $h_i(t)$ with strength $f_i$, whereas the second,
given by $\alpha f_iv_i$, accounts to the coupling
 between the driving strength and velocity $v_i$,  where parameter $\alpha$ quantifies its
 weight.
 Driving forces proportional to the velocity are rarely explored theoretically
 \cite{PhysRevE.88.032102,PhysRevLett.80.5044},
 but they are present in distinct experimental
 studies such as, an electrical force
 stemming from delayed feedback  \cite{dago2022virtual}, self-motile colloidal particles  \cite{howse2007self}, catalytic nanomotors \cite{paxton2004catalytic} and others \cite{schweitzer2003brownian}.

This paper is organized as follows: Sec. II presents the main equations, system thermodynamics and distinct optimization routes. Results and phase diagrams
are presented in Sec. III and conclusions are drawn in Sec. IV.

 \section{Thermodynamics of collisional engines}
One of the simplest engines is composed of a
Brownian particle with  mass $m$  sequentially  placed in contact with a given thermal reservoir and subjected to an external force ${\tilde f}_i(t)$ at each stage.
Each contact  has a duration of $\tau/N$ (with $\tau$ and $N$ being the total time and the number of strokes, respectively) and occurs during the intervals $\tau_{i-1} \le t <\tau_{i}$, where $\tau_{i}=i\tau/N$ for $i=1,..,N$, in which the particle evolves in time according to the following Langevin equation
\begin{equation}
    \frac{d v_i}{dt} =  - \gamma_i v_i +{\tilde f}_i(t)+\xi_{i}(t),
  \label{two_baths_mov1}
\end{equation}
where quantities $v_i$,  $\gamma_i$ and  ${\tilde f}_i(t)$
denote its velocity, the viscous constant and the driving force respectively.
As stated previously, ${\tilde f}_i(t)$ is
given by the  time dependent  driving plus a velocity dependent components${\tilde f}_i(t)=(h_i(t)-\alpha v_{i})f_i$, where $\alpha$ is a constant.
Note that  one recovers the standard collisional engine as $\alpha=0$ \cite{noa2020thermodynamics,noa2021efficient}.
 The  interaction between particle  and the $i$-th environment is described
 by the white-noise stochastic force $\xi_{i}(t)$, satisfying the white-noise properties:
\begin{equation}
 \langle \xi_{i}(t)\rangle=0\quad,\quad\langle\xi_{i}(t)\xi_{i^{\prime}}(t^{\prime}) \rangle= 2\gamma_i T_i \delta_{ii^{\prime}} \delta (t - t^{\prime}),
  \label{two_baths_ruido1}
\end{equation}
  where $T_i$ is the bath temperature.  In 
order to obtain the  thermodynamics, let $P_i(v,t)$  the  velocity probability distribution with time evolution
 described by the Fokker-Planck (FP) equation \cite{mariobook,tome2010,tome2015,broeck2010}:
\begin{equation}
   \frac{\partial P_i}{\partial t} = -f_ih_i(t)\frac{\partial P_{i}}{\partial v}-\frac{\partial J_{i}}{\partial v_{i}} ,  
   \label{64}
\end{equation}
where $J_i$ is given by 
\begin{equation}
  \quad J_i = -\beta_{i}v_{i}P_{i}-\gamma_iT_i\frac{\partial P_{i}}{\partial v_{i}},
  \label{645}
  \end{equation}
  and $\beta_{i}=\gamma_{i}+\alpha f_i$. Note that the  term $\alpha f_i$ can  be incorporated with  $\gamma_{i}v_i$ and can be viewed as a new quantity to be optimized, together the external force $f_i$. For simplifying matters, from now on, we shall assume $\kB=m=1$.

From  the FP equation and
by performing the usual boundary conditions in the space of velocities, 
 in which both $P_{i}(v,t)$ and $J_{i}(v,t)$ vanish as $|v|\rightarrow \infty$, the first and  second law of thermodynamics can be derived.
Starting with the former, the time variation of the energy system $U_i=\langle E_i\rangle$ is given by
 $dU_i/dt = - ( {\dot W_i}+{\dot Q_i})$,
where $\dot W_i(t)$ and ${\dot Q_i}(t)$ denote the work per unity of time (power)
and heat flux from the system to the environment (thermal bath) reading
\begin{equation}
{\dot W_i}(t) = -f_i \langle v_i\rangle(t) \quad {\rm and} \quad {\dot Q_i}(t) = \beta_{i}\langle v_i^{2}\rangle(t)-\gamma_iT_i,
\label{1///12}
\end{equation}
respectively. 
Analogously, time evolution of entropy 
$S_i(t) = - \langle \ln [P_i(v_{i},t)]\rangle$ is given by
\begin{equation}
  \frac{dS_i}{dt}=\frac{1}{\gamma_iT_i}\left[\int_{}^{}\frac{J_{i}^{2}}{P_{i}}dv_{i}+{\beta_{i}}\int_{}^{}v_{i}J_{i}dv_{i}\right],
\end{equation}
where the first and second right terms are identified as the entropy
production rate $\Pi_i(t)$ and entropy flux $-\Phi_i(t)$, respectively \cite{tome2010,tome2015,broeck2010}. The entropy flux can be rewritten in a more convenient way:
\begin{equation}
\label{epfl}
\Phi_i(t) = \frac{\beta_{i}}{\gamma_iT_i}[\beta_{i}\langle v_{i}^{2}\rangle(t)-\gamma_iT_i].
\end{equation}
Summarizing, above expressions for thermodynamic quantities can be calculated
from ensemble averages  $\langle v_i\rangle(t)$ and 
$\langle v_i^2\rangle(t)=b_i(t)+\langle v_i\rangle^2(t)$. Since the coupling 
between velocity and external driving can be incorporated into Eq. (\ref{645}),  the probability distribution has a similar form to the couplingless case \cite{noa2020thermodynamics} and
presents a  Gaussian form: \begin{equation}
P_{i}(v,t) =\frac{1}{\sqrt{2\pi b_i(t)}} \exp{-\frac{1}{2b_i(t)}[v - \langle v_i\rangle(t)]^2}, 
\end{equation}
in which the mean $\langle v_i\rangle(t)$  and variance $b_i(t)= \langle v_i^2\rangle(t) - \langle v_i\rangle^2 (t)$ are time dependent and obey the following
equations
\begin{equation}
    \frac{d \langle v_i\rangle(t)}{dt} =  - \beta_i \langle v_i\rangle(t) +f_ih_i(t)
\end{equation}
and
\beq
    \frac{d b_i(t)}{dt}  = -2 \beta_i b_i(t)  + 2\gamma_i  T_i,
\eeq 
respectively. 
Continuity of $P_{i}(v,t)$  at each stroke implies that
$\langle v_i\rangle(\tau_i)=\langle v_{i+1}\rangle(\tau_i)$ and
$b_i(\tau_i)=b_{i+1}(\tau_i)$ (for all $i=1,...,N$),  
respectively. Since the system returns to the initial
state after a complete period, $\langle v_1\rangle(0)=\langle v_N\rangle(\tau)$ and
$b_1(0)=b_N(\tau)$, 
all averages $\langle v_i\rangle(t)$'s and variances can be solely calculated
in terms of model parameters, that is, from the driving, temperature reservoirs, coupling
$\alpha$ and the period.
By focusing on the simplest design of an engine composed of only two strokes and returning to the initial step after one cycles,
 expressions for averages
and variances can be obtained for an arbitrary driving:
\begin{equation}
    \centering
   \mom{v_1}(t)=e^{-\beta _1 t} \left(f_1 \mathcal{F}_1(t,\alpha)+\frac{f_2
   e^{\frac{\beta _1 \tau }{2}} \mathcal{F}_2(\tau,\alpha )+f_1
   \mathcal{F}_1\left(\frac{\tau }{2},\alpha\right)}{e^{\frac{1}{2}
   \left(\beta _1+\beta _2\right) \tau }-1}\right),
\end{equation}
and
\begin{equation}
    \centering
    \mom{v_2}(t)=e^{-\beta _2 \left(t-\frac{\tau }{2}\right)}5
   \left(f_2 \mathcal{F}_2(t,\alpha)+\frac{f_1 e^{\frac{\beta _2 \tau
   }{2}} \mathcal{F}_1\left(\frac{\tau }{2},\alpha\right)+f_2
   \mathcal{F}_2(\tau,\alpha )}{e^{\frac{1}{2} \left(\beta _1+\beta
   _2\right) \tau }-1}\right),
\end{equation}
for the mean velocities  and
\begin{equation}
    \centering
    b_1(t)=\frac{\gamma  \left(e^{\beta _2 \tau }-1\right) \left(\beta _1
   T_2-\beta _2 T_1\right) e^{\beta _1 \tau -2 \beta _1
   t}}{\beta _1 \beta _2 \left(e^{\left(\beta _1+\beta
   _2\right) \tau }-1\right)}+\frac{\gamma  T_1}{\beta _1},
\end{equation}
and
\begin{equation}
    \centering
    b_2(t)=\frac{\gamma  \left(e^{\beta _1 \tau }-1\right) \left(\beta _2
   T_1-\beta _1 T_2\right) e^{\beta _2 \tau -2 \beta _2
   \left(t-\frac{\tau }{2}\right)}}{\beta _1 \beta _2
   \left(e^{\left(\beta _1+\beta _2\right) \tau
   }-1\right)}+\frac{\gamma  T_2}{\beta _2},
\end{equation}
for variances, respectively,
where $\mathcal{F}_1(t,\alpha)=\int _0^te^{\beta _1 t'}  h_1\left(t'\right)dt'$ and $\mathcal{F}_2(t,\alpha)=\int _{\tau/2}^te^{\beta _2 \left(t'-\frac{\tau}{2}\right)}  h_2\left(t'-\frac{\tau}{2}\right)dt'$.
From above expressions, all thermodynamic quantities are straightforwardly
evaluated. Starting with the work (actually the power), averaged over
a complete period, it  follows that:
\begin{widetext}
\begin{align}
    \overline{\Dot{W}}_1&=-\frac{f_1}{\tau} \int_0^{\frac{\tau }{2}} h_1(t) e^{-(\gamma+\alpha f_1)
   t} \left(f_1 \mathcal{F}_1(t,\alpha)+\frac{f_2 e^{\frac{(\gamma+\alpha f_1) \tau }{2}} \mathcal{F}_2(\tau,\alpha )+f_1
   \mathcal{F}_1\left(\frac{\tau }{2},\alpha\right)}{e^{\frac{1}{2}
   \left[2\gamma+(f_1+f_2)\alpha\right] \tau }-1}\right) 
   dt\\
   \text{and}\nonumber\\
    \overline{\Dot{W}}_2&=-\frac{f_2}{\tau} \int_{\frac{\tau }{2}}^{\tau } h_2(t)
   e^{-(\gamma+\alpha f_2) \left(t-\frac{\tau }{2}\right)}
   \left(f_2 \mathcal{F}_2(t,\alpha)+\frac{f_1 e^{\frac{(\gamma+\alpha f_2) \tau
   }{2}} \mathcal{F}_1\left(\frac{\tau }{2},\alpha\right)+f_2
   \mathcal{F}_2(\tau,\alpha )}{e^{\frac{1}{2} \left[2\gamma+(f_1+f_2)\alpha\right] \tau }-1}\right) \, dt,
   \label{work1}
   \end{align}
   \end{widetext}
respectively, where they were expressed  in terms of forces strengths.  Note that $\overline{\Dot{W}}_i$'s are general and valid for all
components $h_i(t)$'s. In contrast to collisional engines with no velocity
dependent component ($\alpha=0$) \cite{noa2020thermodynamics,noa2021efficient}, they do not present a quadratic dependence on forces $f_1$ and $f_2$, but reduces to
them in such a limit.

Analogously, having   expressions for $\langle v_{1}^{2}\rangle(t)$ and $\langle v_{2}^{2}\rangle(t)$,  averages $\overline{\dot{Q}}_{1}$  ($\overline{\dot{Q}}_{2}$)   can be calculated and decomposed as a sum of two terms: $\overline{\dot{Q}}_{1}=\overline{\dot{Q}}_{1f}+\overline{\dot{Q}}_{f_1,f_2,T_1}$ 
($\overline{\dot{Q}}_{2}=\overline{\dot{Q}}_{2f}+\overline{\dot{Q}}_{f_1,f_2,T_2}$),
the former  term given by $\overline{\dot{Q}}_{1f}=\int_0^{\tau/2}\langle v_{1}\rangle^2(t)dt/\tau$ ($\overline{\dot{Q}}_{2f}=\int_{\tau/2}^{\tau}\langle v_{2}\rangle^2(t)dt/\tau$) and accounting to the contribution
for heat due to drivings, whereas the latter  $\overline{\dot{Q}}_{f_1,f_2,T_1}$  reads $\overline{\dot{Q}}_{f_1,f_2,T_1}= \int_0^{\tau/2} b_1(t)dt/\tau-\frac{\gamma T_1}{2}$,
and describes
the interplay between strength forces $f_i$'s, temperatures of reservoirs $T_i$'s and $\alpha$. From  expressions for $b_i(t)$'s, 
each above component is given by:
\begin{equation}\small
\overline{\dot{Q}}_{f_1,f_2,T_1}=\frac{\left(e^{(1+\alpha f_1) \tau  }-1\right) \left(e^{(1+\alpha f_2) \tau  }-1\right) [\alpha (f_1 T_2-f_2 T_1)+T_2-T_1]}{2
   \tau \left(e^{\tau  (\alpha  (f_1+f_2)+2)}-1\right) (1+\alpha f_1) (1+\alpha f_2) },
   \label{q1f}
\end{equation}
and  $\overline{\dot{Q}}_{f_1,f_2,T_1}=- \overline{\dot{Q}}_{f_1,f_2,T_2}$. Note that their values averaged over a full period implies that
    $\overline{\dot{Q}}_{1}+\overline{\dot{Q}}_{2}+\overline{\dot{W}}_{1}+\overline{\dot{W}}_{2}=0$,
in consistency with  the first law of thermodynamics. Finally,
having $\overline{\dot{Q}}_{i}$'s, the
 the steady entropy production is promptly obtained from Eq. (\ref{epfl}) and given by $\overline{\Dot{Q}_1}/T_1+\overline{\Dot{Q}_2}/T_2$.

\subsection{Constant and linear drivings}
In order to compare such new ingredient with collisional engines \cite{noa2020thermodynamics, noa2021efficient,akasaki2022thermodynamics},
analysis will be exemplified for the two simplest kinds of drivings: 
 constant and linear ones. Both of them have strengths $f_1$ and $f_2$,
the former being time independent with  at $0<t\le \tau/2$ and $\tau/2<t\le \tau$, respectively, whereas the
latter is given by
\begin{equation}
h_i(t)= \left\{\begin{matrix}
 \gamma t;\hspace{1.8cm}0\le t<\tau/2\\\\
\gamma (t-\tau/2);\hspace{0.5cm}\tau/2 \le t<\tau
\end{matrix}\right.
\end{equation}
Thermodynamic quantities are directly evaluated from Eqs. (\ref{work1})-(\ref{q1f}).

\subsection{Efficiency}
In several cases, the entropy production assumes the generic bilinear form $J_lF_l+J_dF_d$.
A common definition of efficiency  in such cases is  given by the ratio between entropy production components
$-J_lF_l/J_dF_d$, describing the partial conversion of    one type of energy,
expressed in terms of a driving force $F_d$ with corresponding flux $J_d$
into another one, characterized by a load force
$F_l$ and flux $J_l$.
Above relation has been used  for describing several systems in nonequilibrium thermodynamics,
such as linear stochastic thermodynamics \cite{proesmans2015efficiency,proesmans2016brownian,PhysRevLett.95.190602}, systems in contact
presenting a single worksource and heatsource \cite{gatien}, work-to-work transducers \cite{herpich,herpich2,busiello2022hyperaccurate,liepelt1,liepelt2} and others.
On the other hand, the class of engines we are investigating can be associated  with three thermodynamic forces (two of them are related to  $f_1$, $f_2$ and the third with the difference of temperatures), implying the usage of above ratio as a
dubious measure of the system performance for $T_1\neq T_2$. For this reason, we consider
a  definition of efficiency given by \cite{noa2021efficient,mamede2021obtaining,akasaki2022thermodynamics}
\begin{equation}\label{eta} 
    \eta=-\frac{\cal P}{\overline{\dot{W}}_{1}+\overline{\dot{Q}}_{1}\Theta(-\overline{\dot{Q}}_{1})+\overline{\dot{Q}}_{2}\Theta(-\overline{\dot{Q}}_{2})},
\end{equation}
also expressing the partial conversion of a given amount of  energy under the form of input heat $\overline{\dot{Q}}_{1}\Theta(-\overline{\dot{Q}}_{1})+\overline{\dot{Q}}_{2}\Theta(-\overline{\dot{Q}}_{2})<0$  ($\Theta(x)$ being the Heaviside function) plus input work $\overline{\dot{W}}_{1}<0$   into  power output ${\cal P}\equiv \overline{\dot{W}}_{2}\ge 0$. Eq. (\ref{eta}) reduces to the previous definition for $\Delta T=T_1-T_2=0$
in which  output and input works are related to fluxes as
 ${\cal P}=-TJ_lF_l$ and $\overline{\dot{W}}_{1}=-TJ_dF_d$   \cite{proesmans2016brownian,noa2020thermodynamics,mamede2021obtaining,akasaki2022thermodynamics}.
Since realistic engines operate at finite time, we are going
to exploit distinct routes for optimizing the system performance for finite 
$\tau$: by maximizing $\eta$ and ${\cal P}$ with
respect to the $f_2/f_1$ and $\alpha$. Although such maximizations
can be directly performed from the expressions for ${\cal P}$ and $\eta$ from Eqs. (\ref{work1}) and  (\ref{eta}), respectively, these expressions are little instructive, due to the complex interplay between
$\alpha$ and $f_2$. For this reason, in the next section, we shall present
distinct approaches/reasonings for obtaining some maximized quantities with respect to $\alpha$ (for fixed
$f_1$ and $f_2$) and optimized $f_2$ (for fixed $\alpha$ and $f_1$).

\subsection{Linear approximation  for the  power output}

As stated before, the role of coupling $\alpha$ and its interplay with $f_i$'s and $T_i$'s is not evident. In order to
obtain some insight about it, the analysis is performed for small $\alpha$, in which 
the average ${\cal P}$  is decomposed  in the following way:
\begin{equation}\label{W12}
  {\cal P}\approx {\cal P}_{0}+\alpha{\cal P}_{\alpha}.
\end{equation}
where ${\cal P}_{0}$ accounts the average power (calculated at the $2$-th stage) for the couplingless case. According to Refs. \cite{noa2021efficient,akasaki2022thermodynamics}, ${\cal P}_{0}$ can be expressed
in terms of Onsager coefficients,
${\cal P}_{0}=-T_2(L_{21}f_1+L_{22}f_2)$, where each coefficient $L_{2i}$ is given by
\begin{widetext}
  \begin{equation}
    \begin{split}
L_{22}&=\frac{m}{T_2\tau  \left(e^{\gamma  \tau }-1\right)}\left [\int_{\tau/2}^{\tau } h_2(t) e^{-\gamma  t} \, dt \int_{\tau/2}^{\tau } h_2(t^\prime) e^{\gamma  t^\prime} \, dt^\prime+\left(e^{\gamma  \tau }-1\right) \int_{\tau/2}^{\tau } h_2(t) e^{-\gamma  t}   \int_{\tau/2}^t h_2(t^\prime) e^{\gamma  t^\prime} \, dt^\prime dt \,\right ],\\
  L_{21}&=\frac{m e^{\gamma  \tau } }{T_2\tau  \left(e^{\gamma  \tau }-1\right)} \int_0^{\tau/2} h_1(t^\prime) e^{\gamma  t^\prime} \, dt^\prime \int_{\tau/2}^{\tau } h_2(t) e^{-\gamma  t} \, dt.
  \label{l2ew}
\end{split}
\end{equation}
\end{widetext}
In a similar fashion, expressions for the average work in the first stage assumes the form $\overline{\dot{W}}_{10}=-T_1(L_{11}f_1+L_{12}f_2)$, also
expressed in terms of Onsager coefficients $L_{1i}$'s given by
 \begin{widetext}
  \begin{equation}
    \begin{split}
   L_{11}&= \frac{m}{T_1\tau  \left(e^{\gamma  \tau }-1\right)}  \left[\left(e^{\gamma  \tau}-1\right) \int_0^{\tau/2}h_1(t) e^{-\gamma  t}\int_0^t  h_1(t^\prime) e^{\gamma t^\prime} \, dt^\prime dt \, +\int_0^{\tau/2} h_1(t) e^{-\gamma t}dt \,  \int_0^{\tau/2} h_1(t^\prime) e^{\gamma  t^\prime} \, dt^\prime\right ],\\
  L_{12}&=\frac{m}{T_1\tau  \left(e^{\gamma  \tau }-1\right)} \int_0^{\tau/2} h_1(t) e^{-\gamma t} \, dt \int_{\tau/2}^{\tau } h_2(t^\prime) e^{\gamma  t^\prime} \, dt^\prime.\\
  \label{l1ew}
    \end{split}
\end{equation}
\end{widetext}
Reciprocal  relations for
  cross coefficients $L_{12}$ and $L_{21}$ are derived when drivings are reversed and  indices $1\leftrightarrow 2$ exchanged \cite{rosas2,akasaki2022thermodynamics}.
The linear contribution
${\cal P}_{\alpha}$ can also be expressed in the following generic form:
\begin{equation}\label{W12}
    {\cal P}_{\alpha}=T_2f_2(\widetilde{L}_{222}f_2^{2}+\widetilde{L}_{211}f_1^2+\widetilde{L}_{221}f_2f_1),
\end{equation}
 where general expressions for coefficients ${\tilde L}_{2jk}$'s are listed in Appendix \ref{appa} and \ref{appb} for generic and constant and linear drivings, respectively.  However,
 in contrast to Onsager  ones, above coefficients $\widetilde{L}_{ijk}$'s do not necessarily satisfy the standard reciprocal relations \cite{noa2021efficient,akasaki2022thermodynamics,rosas2}.  
Analogously to ${\cal P}_\alpha$,  the linear contribution for $\overline{\dot{W}}_{1\alpha}$  can also be expressed
in following form given by
$\widetilde{\overline{\dot{W}}}_{1\alpha}=T_1f_1(\widetilde{L}_{111}f_1^{2}+\widetilde{L}_{221}f_2^2+\widetilde{L}_{112}f_2f_1)$, whose coefficients $\widetilde{L}_{1jk}$'s are also listed in Appendix \ref{appa}. 
From Eq. (\ref{W12}), the optimal
force $f_{2P}$ providing maximum power  ${\cal P}_{P}$
are straightforwardly obtained and given by
\begin{equation}
 f_{2P}=\frac{-(L_{22}+\alpha  f_1 \tilde {L}_{221})}{3\alpha \tilde{L}_{222}}\left[1- \sqrt{1-\frac {3\alpha f_1 \tilde {L}_{222}(L_{21}+\alpha  f_1 \tilde {L}_{221})}{(L_{22}+\alpha  f_1 \tilde {L}_{221})^2}}\right],
 \end{equation}
 and
  \begin{widetext}
 \begin{align}
    \frac{ \mathcal{P}_{P}}{T_2}&=\frac{2 \left(\alpha  f_1 \tilde{L}_{221}+L_{22}\right){}^2 \left(\alpha  f_1 \tilde{L}_{221}+L_{22}-\mathcal{A}\right)-3 \alpha  f_1 \tilde{L}_{222}
   \left(\alpha  f_1 \tilde{L}_{211}+L_{21}\right) \left(3 \alpha  f_1 \tilde{L}_{221}+3 L_{22}-2 \mathcal{A}\right)}{27 \alpha ^2 \tilde{L}_{222}^2}\nonumber, 
 \end{align}
\end{widetext}
 respectively, where   $\mathcal{A}$ reads
   $\mathcal{A}=\sqrt{\left(\alpha  f_1 \tilde{L}_{221}+L_{22}\right){}^2-3 \alpha  f_1 \tilde{L}_{222} \left(\alpha  f_1 \tilde{L}_{211}+L_{21}\right)}$. Note that one recovers the expression $2f_{2P}=-L_{21}/L_{22}$
  and ${\cal P}_{P}=T_2L_{21}^2f_1/4L_{22}$ as $\alpha=0$.

\subsection{Approximate descriptions for maximum efficiencies}\label{secd}
 Since the average heat components $\overline{\dot{Q}}_{1f}$ and $\overline{\dot{Q}}_{2f}$ are always positive, the system solely will receive heat from the $i-$th thermal
bath from a   temperature difference $\Delta T$ in which $\overline{\dot{Q}}_{i}=\overline{\dot{Q}}_{if}+\overline{\dot{Q}}_{f_1,f_2,T_i}<0$.
Giving that above condition is always fulfilled for  large $\Delta T$ and by the fact that the power output ${\cal P}$ does not depend on the temperatures, the efficiency
of thermal engines  for
$\alpha=0$ always decreases when compared to its corresponding work-to-work converter  $\eta_{wtw}=-{\cal P}/\overline{\dot{W}}_{1}$.  However, a coupling between drivings and velocities  makes 
 possible to (properly) adjust the coupling  ensuring a maximum efficiency.
Despite 
the complex interplay between $\alpha$ and $f_1,f_2$ leads to very cumbersome expressions for
maximized quantities (above all the efficiency), it is possible to predict optimized
expressions for efficiency by means of two simple reasonings, as described as follows:
The first analysis can be performed in the regime of small $\alpha$,
 in which, in  similarity to the expansion for ${\cal P}$ and $\overline{\dot{W}}_{1}$,
one assumes the following expansions for $\overline{\dot{Q}}_{1}$ and
$\overline{\dot{Q}}_{2}$:
\begin{equation}
    \overline{\dot{Q}}_{1}=\overline{\dot{Q}}_{10}+\alpha\widetilde{\overline{\dot{Q}}}_{1\alpha}\quad,\quad\overline{\dot{Q}}_{2}=\overline{\dot{Q}}_{20}+\alpha\widetilde{\overline{\dot{Q}}}_{2\alpha}, 
\end{equation}
By inserting above expressions in Eq. (\ref{eta}) and considering up to the linear term,
the efficiency is given by $\eta \approx \eta_0+\alpha \eta_\alpha$, where $\eta_0$ and $\eta_\alpha$
read:
\begin{equation}\label{ETA0}
     \eta_{0}=-\frac{{\cal P}_0}{\overline{\dot{W}}_{10}+\overline{\dot{Q}}_{10}\Theta\left(-\overline{\dot{Q}}_{10}\right)+\overline{\dot{Q}}_{20}\Theta\left(-\overline{\dot{Q}}_{20}\right)},
\end{equation}
and
\begin{equation}\label{ETA1}
     \eta_{\alpha}=\frac{{\cal P}_\alpha-\eta_{0}\left[\overline{\dot{W}}_{1\alpha}+ \overline{\dot{Q}}_{1\alpha}\Theta\left(-\overline{\dot{Q}}_{10}\right)+\overline{\dot{Q}}_{2\alpha}\Theta\left(-\overline{\dot{Q}}_{20}\right) \right]}{\overline{\dot{W}}_{10}+\overline{\dot{Q}}_{10}\Theta\left(-\overline{\dot{Q}}_{10}\right)+\overline{\dot{Q}}_{20}\Theta\left(-\overline{\dot{Q}}_{20}\right)},
\end{equation}
respectively. Note that $\eta_0$ solely depends on $0$-th order quantities (as expected), whereas  $\eta_\alpha$ depends on $\eta_0,\overline{\dot{W}}_{i\alpha}$'s and $\overline{\dot{Q}}_{i\alpha}$'s. Maximization of $\eta$ with respect to $f_2$, providing $f_{2mE}$, can be calculated from  Eqs. (\ref{ETA0}) and (\ref{ETA1}).


Contrariwise, for  the case in which above approximation for small $\alpha$
is not valid, maximization of efficiency can be carried out
by means of  a simple argument, as described as follows:
Let us consider the case in which the average heat component  $\overline{\dot{Q}}_{f_1,f_2,T_i}$ dominates over  $\overline{\dot{Q}}_{if}$  ($|\overline{\dot{Q}}_{f_1,f_2,T_i}|\gg\overline{\dot{Q}}_{if}$). Although
this is verified for  sufficient large $|\Delta T|$ and  fixed $f_1,f_2$, 
such condition can be fulfilled for other interplay among parameters. For situations
in which  the power output monotonically increases (this is promptly verified for
the drivings considered here) upon $\alpha$ is varied, 
the efficiency can be enhanced by searching for the optimal coupling $\alpha_{E}$,   in which $\overline{\dot{Q}}_{if}+\overline{\dot{Q}}_{f_1,f_2,T_i} \approx \overline{\dot{Q}}_{f_1,f_2,T_i}=0$:
 \begin{equation}
 \alpha_{E}=\frac{T_1-T_2}{f_1 T_2-f_2 T_1}. 
 \label{alo}
 \end{equation}
Note that above approximate  relation is general and expresses the interplay among driving strengths
 and temperatures and approaches to $0$
as $\Delta T \rightarrow 0$ for finite $f_1-f_2$, showing that forces proportional
to the velocity can increase the efficiency for suited choice of temperatures and forces.
The corresponding maximum efficiency $\eta_{\alpha_{E},f_1,f_2,\Delta T}$ reduces to the work-to-work converter expression given by
\begin{equation}
{\bar \eta}_{f_2,f_1,\Delta T}=-\frac{{\cal P}^*_{E}}{\overline{\dot{W}}_{1E}^*},
\label{maxe}
\end{equation}
with   ${\cal P}^*_{E}={\cal P}(f_1,f_2,\Delta T,\tau)$ and $\overline{\dot{W}}_{1E}^*=\overline{\dot{W}}_{1}(f_1,f_2,\Delta T,\tau)$ denoting the  ${\cal P}$ and $\overline{\dot{W}}_{1}$
   evaluated at $\alpha=\alpha_{E}$, respectively. Note that   Eq. (\ref{maxe}) 
   solely depends on $f_1,f_2,\tau$ and $\Delta T$. The efficiency can also be  maximized
   with respect to  $f_2$ (for fixed $\Delta T$) or $\Delta T$ (for fixed $f_2)$.
Although it can be directly carried out by a simultaneous maximization of Eq. (\ref{eta}), an approximate  expression for
the simultaneous maximum efficiency $\eta^*_{f_1,\theta}$ ($\theta=f_2$ or $\Delta T$) is obtained by searching for $f_2$ or $\Delta T$ that maximizes Eq. (\ref{maxe}):
\begin{equation}
\eta^*_{f_1,\theta}=-\frac{{\cal P}^*_{mE}}{\overline{\dot{W}}^*_{1mE}},
\label{maxeg}  
\end{equation}
where  ${\cal P}^*_{mE}={\cal P}(f_1,\theta_E,\tau)$ and $\overline{\dot{W}}_{1mE}^*=\overline{\dot{W}}_{1}(f_1,\theta_E,\tau)$
denoting the  ${\cal P}$ and $\overline{\dot{W}}_{1}$
   evaluated at $\alpha=\alpha_{E}$ and $f_2=f_{2E}$ ( $\theta=\Delta T$) or $\Delta T=\Delta T_E$
   ($\theta=f_2$)
   respectively.

\section{Results}
In all cases, analysis will be carried out  for constant and linear
drivings and the following parameters choices $m=\tau=\gamma=f_1=1$. Expressions for Onsager and  coefficients from the linear analysis
are listed in Appendix \ref{appb}.
\begin{figure}
     \centering
     \includegraphics[scale=0.75]{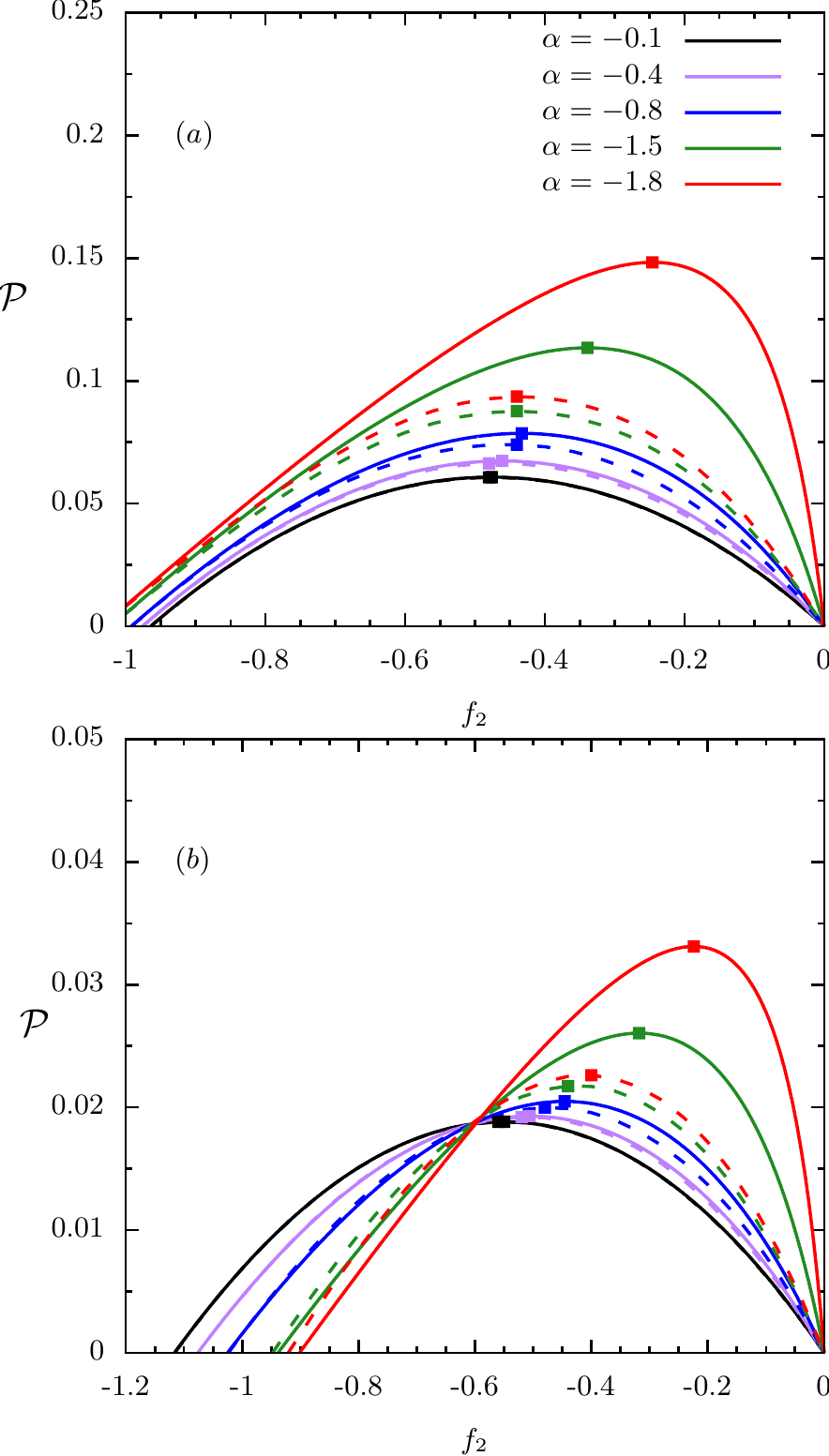}
     \caption{For constant $(a)$ and linear $(b)$ drivings, the depiction of power output $\mathcal{P}$ (continuous) and
     those from  a linear analysis (dashed), where squares correspond to $f_{2P}$'s ensuring maximum power   $\mathcal{P}_{P}$'s. Parameters: $m=\tau=\gamma=1$, $f_1=1$.}
     \label{fig2}
 \end{figure} 
In the first round of analysis, the influence of $\alpha$ over the power output ${\cal P}$ and efficiency $\eta$ are exemplified for some sort of parameters, as
depicted in Figs. \ref{fig2}-\ref{fig3}  for constant and linear drivings.
${\cal P}$ (Fig. \ref{fig2})   monotonically increases with 
the absolute value of $\alpha$, having this feature captured by the linear analysis for small $\alpha$. In other words, apart from the increase of power as the absolute value of $\alpha$ increases, there is no optimal coupling leading to maximum power, implying 
its maximization solely with respect to the force $f_2$ in which  ${\cal P}_P$.  
 
The   influence of $\alpha$  over the efficiency 
is more revealing and exemplified in Fig. \ref{fig3} for two representative
temperature differences: $\Delta T=0.25$ $(T_1>T_2)$ and $\Delta T=-0.25$ $(T_1<T_2)$.
\begin{figure}
    \centering
\includegraphics[scale=0.6]{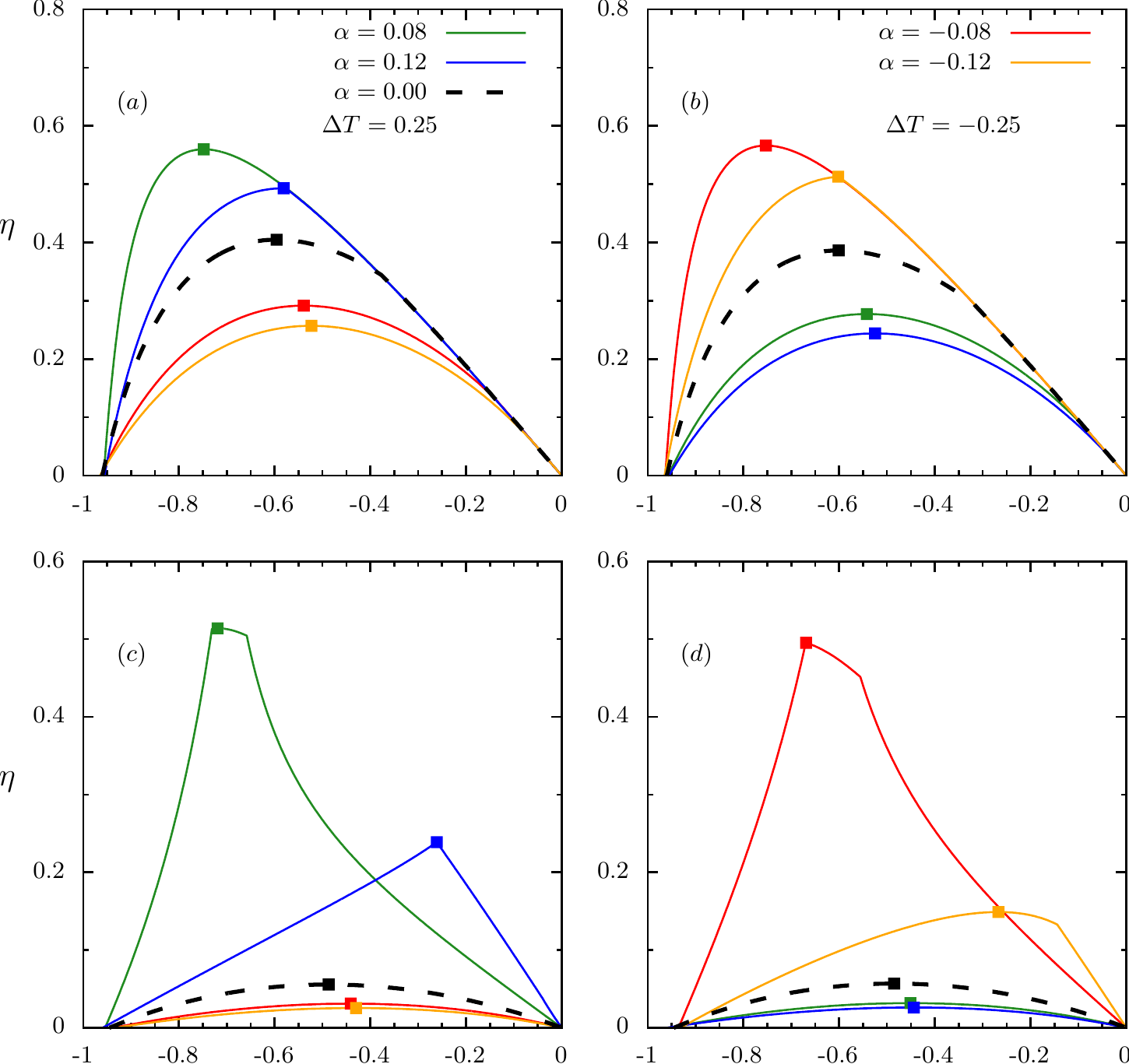}
    \caption{For constant drivings, the depiction of efficiency $\eta$ for representative
    values of $\alpha$ and distinct $\Delta T$'s.
 In $(a)/(c)$ and $(b)/(d)$, $T_1$ reads
 $T_{1}=2.0$ and $T_{1}=1.75$, respectively. Squares denote the associate maximum efficiencies (with respect to the $f_2$).  Parameters: $m=\tau=\gamma=f_2=1$. }
    \label{fig3}
\end{figure}
In both cases,  efficiencies are rather small when  $\alpha=0$ (couplingless case) and an optimal coupling between driving and velocities   ensures  substantial increases (see e.g. dashed lines). Also,
 efficiency curves behave quite differently  with respect to $\alpha=0$.  This is
 due to the influence of parameters (mainly $f_2, \alpha$ and $\Delta T$) 
 on the average works and  on the amount of a received heat  [see e.g. denominator from Eq. (\ref{eta})] and it is more significant for linear drivings (see e.g. Fig. \ref{fig3}), where efficiencies
are just smaller \cite{noa2020thermodynamics,noa2021efficient}.  
 Thus, whenever 
 ${\cal P}$ ($\overline{\dot{W}}_{1}^*$) always increases with the absolute
 value of $\alpha$, there is an optimal coupling  $\alpha_E$ 
controlling/decreases the amount of "wasted" average heat. In particular, the optimal coupling
$\alpha_E$ is positive and negative for $T_1>T_2$ and $T_1<T_2$, respectively.

\begin{figure}
     \centering
     \includegraphics[scale=0.7]{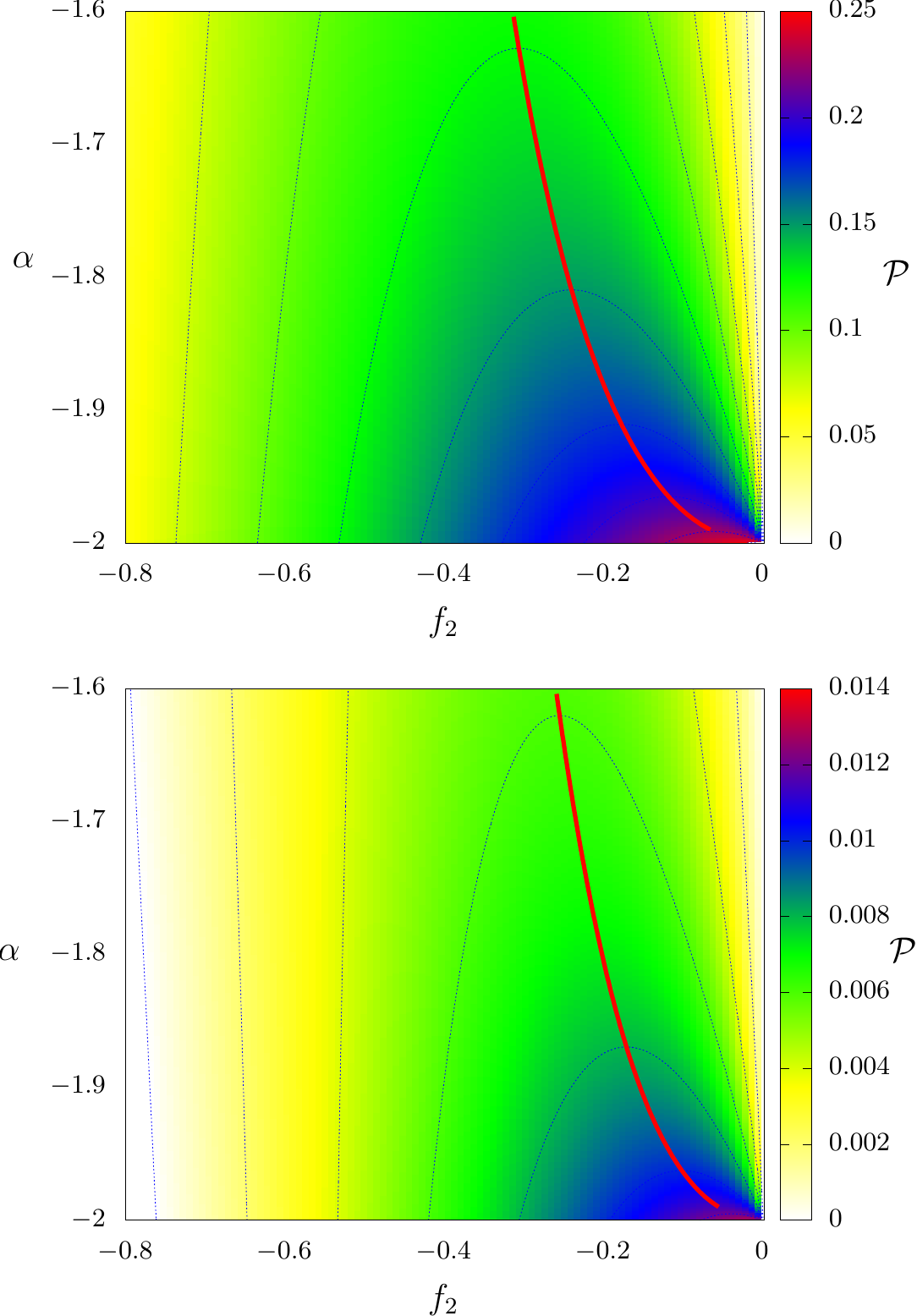}
     \caption{For constant (top) and linear (bottom) drivings, the depiction of power output $\mathcal{P}$ in the plane $(f_{2},\alpha)$. For each $\alpha$, red lines 
     denote the locus ($f_{2P},{\cal P}_{P}$) of maximum  ${\cal P}$ with respect to $f_2$.  Parameters: $m=\tau=\gamma=1$, $f_1=1$.\textcolor{red}.}
     \label{fig22}
 \end{figure}

\begin{figure*}
  \centering
\includegraphics[scale=0.7]{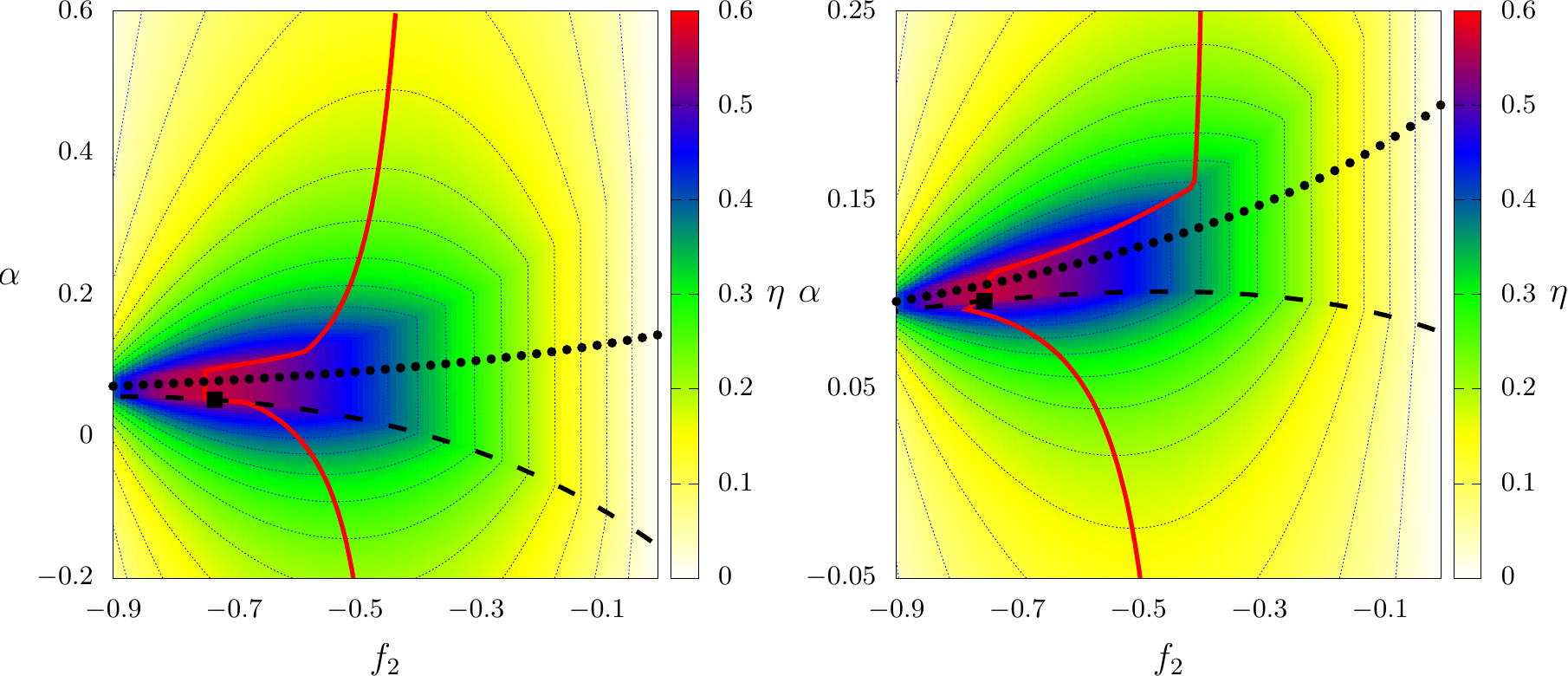}
\caption{
For the same parameters of Fig. \ref{fig3}, the efficiency phase diagrams $\alpha$ versus $f_2$ for  $\Delta T=0.25$ (left) and $-0.25$ (right). Continuous,
 dashed and dotted 
 lines correspond to the maximization with respect to $f_2$ (fixed $\alpha$), $\alpha$ (fixed $f_2$) and approximate  [from Eq. (\ref{alo})], respectively. "Squares" denote the simultaneous maximization with
respect to $\alpha$ and $f_2$.}
\label{fig5} 
\end{figure*}

\begin{figure*}
  \centering
\includegraphics[scale=0.7]{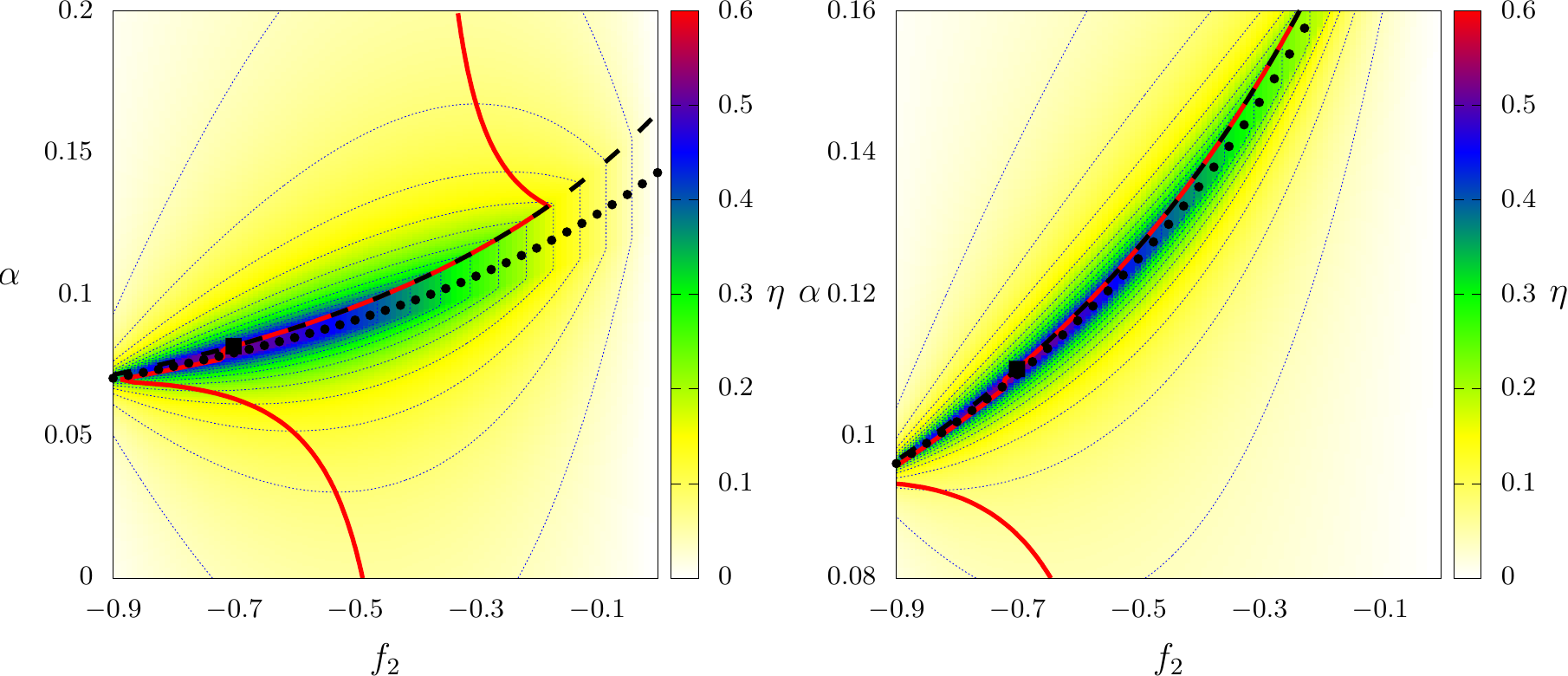}
\caption{For the same parameters of Fig. \ref{fig3}, the efficiency phase diagrams $\alpha$ versus $f_2$ for  $\Delta T=0.25$ (left) and $-0.25$ (right). Continuous,
 dashed and dotted 
 lines correspond to the maximization with respect to $f_2$ (fixed $\alpha$), $\alpha$ (fixed $f_2$) and approximate  [from Eq. (\ref{alo})], respectively. "Squares" denote the simultaneous maximization with
respect to $\alpha$ and $f_2$.}
\label{fig6}
\end{figure*}

\begin{figure}
    \centering
    \includegraphics[scale=0.75]{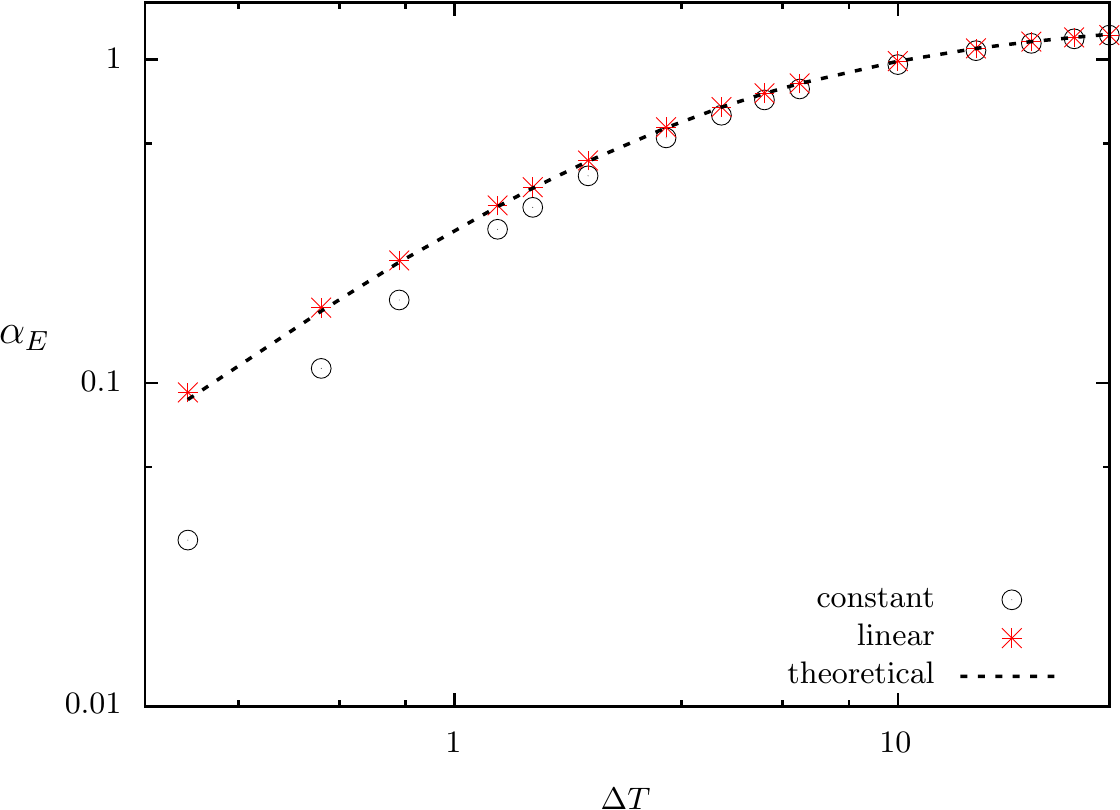}
    \caption{Comparison among logarithm plots of $\alpha_{E}$ obtained from  Eq.(\ref{alo}) (dashed lines) and from direct maximization of Eq. (\ref{eta})
    for constant (circles) and linear (stars) drivings. Parameters: $T_2=1.5$, $m=\tau=\gamma=1,f_1=1$ and $f_2=-0.75$.}
    \label{fig2a}
\end{figure}

For above set of parameters,  Figs. \ref{fig22}-\ref{fig6}  provides 
a global overview about the role of $\alpha$ by depicting 
 heat maps (phase diagrams) for ${\cal P}$ and $\eta$ for linear and constant drivings.  
As in Fig. \ref{fig2}, ${\cal P}$ monotonically increases with the coupling
and providing, for all values of
$\alpha$, optimal $f_{2P}$'s ensuring maximal ${\cal P}_{P}$
(red lines). 
 Contrasting to the power output, in which a simultaneous maximization of power is not possible,
 efficiencies phase diagrams (Figs. \ref{fig5} and \ref{fig6}) exhibit a central region in which efficiency can be simultaneously maximized.  Maximum lines behave differently, reflecting
 the distinct dependence between $\eta$ with $\alpha$ and $f_2$. 
 They meet at  the vicinity of global maximum.  Approximate curves (dotted lines), obtained
 from Eq. (\ref{alo}),  approach  to exact ones (dashed)  as $\Delta T$ is
 raised. They are always closer to each other 
 for linear than constant drivings. Such findings are complemented in Fig. \ref{fig2a}, consistent  with reliability by neglecting  $\overline{\dot{Q}}_{if}$ as $|\Delta T|$ raises.

\begin{figure}
    \centering
    \includegraphics[scale=0.7]{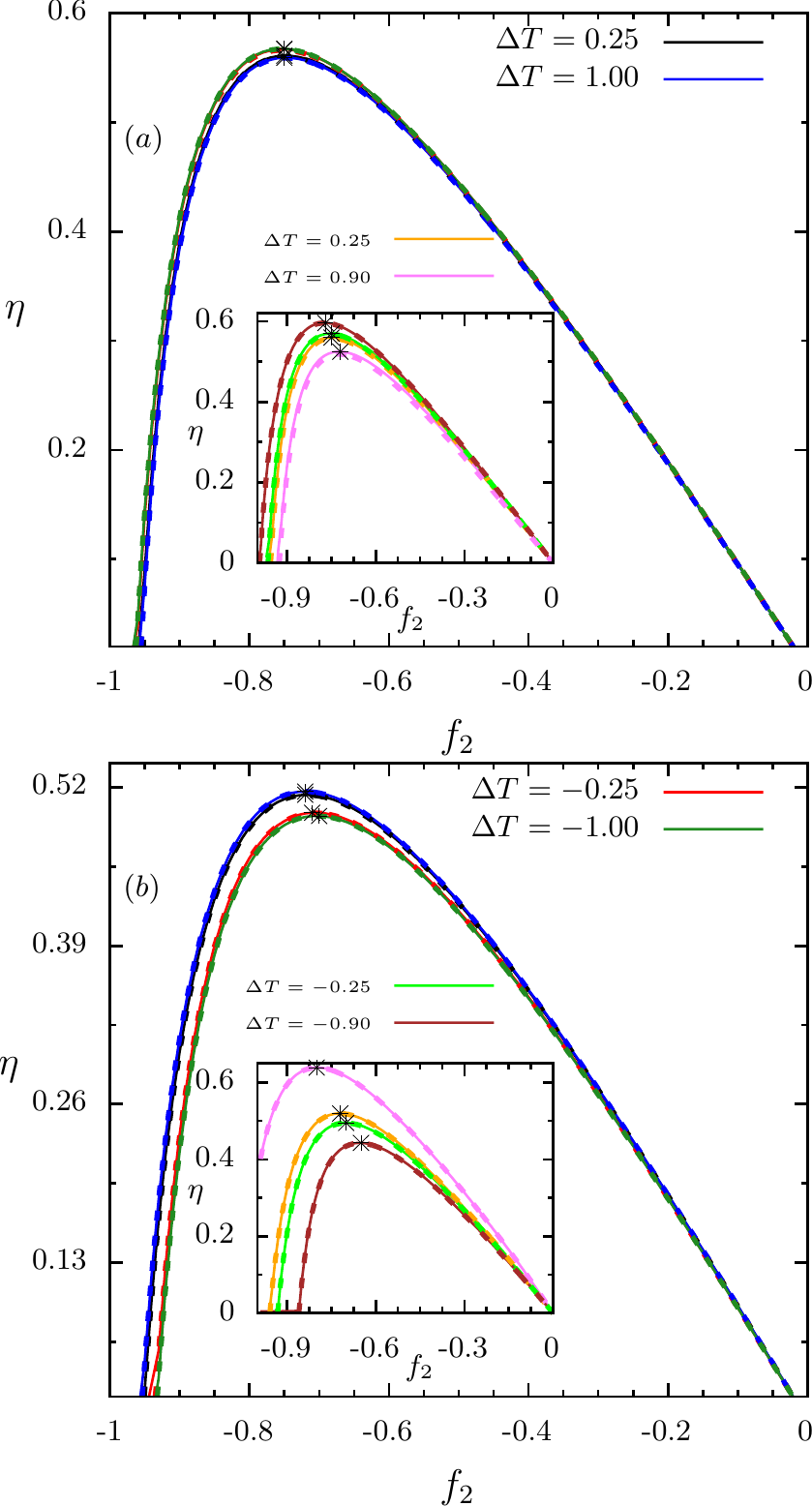}
    \caption{For the same parameters from Figs. \ref{fig5} and \ref{fig6},  main panels  show efficiency maxima
    ${\bar \eta}_{f_2,f_1,\Delta T}$ versus $f_2$ for distinct $\Delta T$'s
    for constant and linear drivings, respectively. Inset: Results
    for the same $\Delta T$
    but for $T_1=0.1,1,1$ and $1.25$, respectively. Stars denote the prediction
from Eq. (\ref{maxeg}) for
    maximized efficiencies
    ${\bar \eta}_{f_2,f_1,\Delta T}$'s with respect to $\alpha$ and $f_2$. Continuous and dashed lines correspond to exact and efficiencies evaluated from Eq. (\ref{maxe}), respectively.}
    \label{max}
\end{figure}
Fig. \ref{max} shows maximum efficiencies  
${\bar \eta}_{f_2,f_1,\Delta T}$ and $\eta^*_{f_1,\Delta T}$, obtained
from direct maximization and by comparing expressions from Eqs.  (\ref{maxe}) and (\ref{maxeg}), respectively. Note the excellent agreement
between exact and approximate expressions (deviations  among curves are almost imperceptible), reinforcing the search for optimal parameters for maximum efficiencies. 
At the vicinity of optimal couplings, they are substantially larger than ${\bar \eta}_{f_1,\Delta T}=0.4049/0.0555$ and $0.1822/0.0151$ (see e.g. Fig \ref{fig3}), obtained
for the couplingless constant/linear cases for $\Delta T=0.25$ and $1$, respectively. 

\begin{figure}
    \centering
    \includegraphics[scale=0.8]{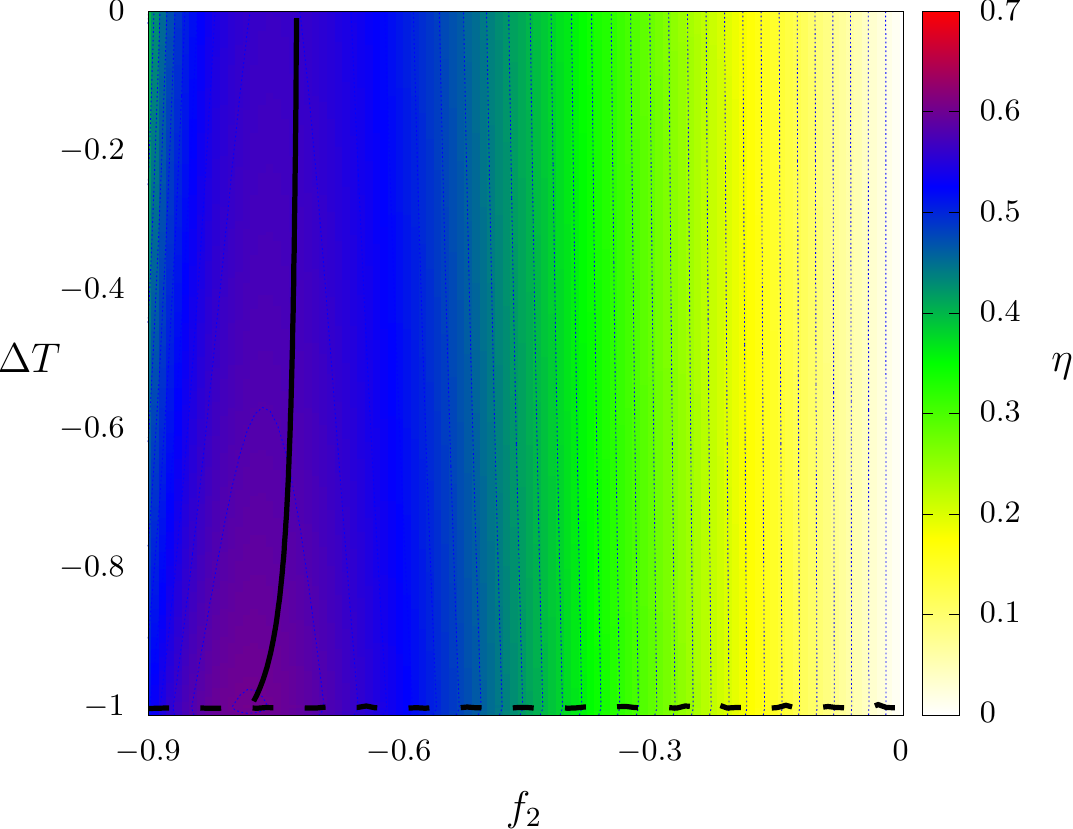}
    \caption{For constant drivings, the efficiency phase diagram $\Delta T$ versus $f_2$. Continuous and dashed lines  denote the maximization respect to the  $f_2$ and $\Delta T$, respectively.}
    \label{maxd}
\end{figure}

Finally, Fig. \ref{maxd} illustrates a phase diagram $\Delta T$ versus
$f_2$ for constant drivings. Note that  the increase  of   $\Delta T$  together
an optimal choice of $f_2$ increases the efficiency. Thus, the coupling  may also provide a suitable choice of temperature difference in order to enhance the efficiency or  furnish a desirable value for it.

  \section{Conclusions}
  Collisional  Brownian engines constitute a very simple class of machines having
  thermodynamic properties exactly obtained irrespective the driving, temperature of thermal
  baths and the duration of each stage. Notwithstanding, its performance can decrease
substantially depending on the way it is projected (period, duration of stage, temperature of baths and drivings).  In order to address possible
improvements in such class of systems,  the influence of velocity  driving component
  was  introduced and analyzed from the framework of stochastic thermodynamics. Results for constant and linear drivings reveal that the it  can be conveniently considered in order to optimize efficiency, even for large temperature differences between thermal reservoirs, where
the couplingless engine operates very inefficiently.  Distinct maximization
routes  were considered and  substantial improvements case can be gained. Despite results  the
absence of a simultaneous maximization for the power output  for
constant and linear drivings, we underscore a reliable choice of coupling
$\alpha$ can be  taken for ensuring a  compromise
between the power output  and efficiency.

As  potential perspectives of the present work, it might be interesting
to address  other kinds of maximizations, such as by holding
the dissipation fixed as well as the  influence  of the coupling
in such cases. Finally, it might also be
remarkable to extend the collisional approach for massive
Brownian particles in order to compare their performances.
\section{Acknowledgment}
I. N. Mamede and C. E. F. acknowledge the financial support from FAPESP under
grants  2021/12551-8 and 2021/03372-2, respectively. The financial supports from CNPq and CAPES are also acknowledged.

\bibliographystyle{apsrev4-1}

\bibliography{refs}
 
 \onecolumngrid
\begin{center}
\textbf{\large Appendix}
\end{center}
\setcounter{equation}{0}
\setcounter{figure}{0}
\setcounter{table}{0}
\setcounter{page}{1}
\setcounter{section}{0}
\makeatletter
\renewcommand{\theequation}{A\arabic{equation}}
\renewcommand{\thefigure}{A\arabic{figure}}
\renewcommand{\citenumfont}[1]{A#1}
 
 \subsection{Coefficients of the linear approximation for small couplings }\label{appa}
 Below, we list  the expressions for coefficients $\tilde{L}_{ijk}$'s
 from the linear expansion of ${\cal P} $ and $\overline{\dot{W}}_{1}$ for generic drivings $h_1(t)$ and $h_2(t)$:
 




\begin{equation}
    \centering
    \tilde{L}_{111}=\frac{\alpha}{2\tau T_1}{\int_0^{\frac{\tau }{2}}e^{-\gamma t} 
\left[\frac{\mathcal{F}_1\left(\frac{\tau }{2},0\right) \left(e^{\gamma \tau } (2 t+\tau )-2 t\right)}{\left(e^{\gamma  \tau }-1\right)^2}+2t \mathcal{F}_1(t,0)\right] \, dt},
\label{l111}
\end{equation}

\begin{equation}
    \centering
    \tilde{L}_{112}=\frac{\alpha  e^{\frac{3 \gamma  \tau }{4}} \sinh \left(\frac{\gamma\tau }{4}\right)}{\gamma ^2 \tau\left(e^{\gamma  \tau }-1\right)^2T_1}{ \left[\gamma  \tau  \mathcal{F}_1\left(\frac{\tau }{2},0\right)+\mathcal{F}_2(\tau ,0) \left(4 \sinh \left(\frac{\gamma  \tau \
}{2}\right)-\gamma  \tau \right)\right]},
\label{l112}
\end{equation}

\begin{equation}
    \centering
    \tilde{L}_{122}=\frac{\alpha  e^{\gamma  \tau } \left(e^{\frac{\gamma  \tau }{2}}-1\right) \mathcal{F}_2(\tau ,0)}{2 \gamma  \left(e^{\gamma  \tau }-1\right)^2T_1},
    \label{l122}
\end{equation}

\begin{equation}
    \centering
    \tilde{L}_{222}=\frac{\alpha  }{2 \tau\left(e^{\gamma  \tau }-1\right)^2T_2}{\int_{\frac{\tau }{2}}^{\tau } {e^{\frac{1}{2} \gamma  (\tau -2 t)} \left[\mathcal{F}_2(\tau ,0) \left(2 t \left(e^{\gamma  \tau }-1\right)+\tau \right)-\left(e^{\gamma  \tau}-1\right)^2 (\tau -2 t) \mathcal{F}_2(t,0)\right]} \, dt},
    \label{l222}
\end{equation}

\begin{equation}
    \centering
    \tilde{L}_{221}=\frac{\alpha  \left[\coth \left(\frac{\gamma  \tau }{4}\right)-1\right] \text{sech}^2\left(\frac{\gamma  \tau }{4}\right) \left[\mathcal{F}_1\left(\frac{\tau }{2},0\right) \left(4 \sinh \left(\frac{\gamma  \tau }{2}\right)-\gamma  \tau \right)+\gamma  \tau  \mathcal{F}_2(\tau ,0)\right]}{16 \gamma ^2 \tau T_2},
    \label{l221}
\end{equation}
and
\begin{equation}
    \centering
    \tilde{L}_{211}=\frac{\alpha  e^{\gamma  \tau } \left(e^{\frac{\gamma  \tau }{2}}-1\right) \mathcal{F}_1\left(\frac{\tau }{2},0\right)}{2 \gamma  \left(e^{\gamma  \tau }-1\right)^2T_2}.
    \label{l211}
\end{equation}

\subsection{Coefficients of the linear approximation for constant and linear drivings } \label{appb}
 
 As stated in the main text, by inserting explicit
 expressions for drivings $h_1(t)$ and $h_2(t)$, Onsager
 coefficients $L_{ij}$'s and $\tilde{L}_{ijk}$'s 
 can be straightforwardly obtained from Eqs. (\ref{l2ew})-(\ref{l1ew}) and (\ref{l111})-(\ref{l211}), respectively.
 For constant  drivings,  we arrive at the following expressions:
 \begin{equation}
     T_1L_{11}=T_2L_{22}=\frac{1}{\gamma}\left(\frac{1}{2}-\frac{\tanh \left(\frac{\gamma  \tau }{4}\right)}{\gamma\tau}\right)\quad,\quad T_1L_{12}=T_2L_{21}=\frac{1}{\gamma ^2 \tau }\tanh \left(\frac{\gamma  \tau }{4}\right) 
 \end{equation}
for  Onsager ones and

\begin{equation}
    T_1\tilde{L}_{111}=T_2\tilde{L}_{222}=\frac{\gamma  \tau  \left(\text{sech}^2\left(\frac{\gamma  \tau }{4}\right)+4\right)-16  \tanh \left(\frac{\gamma  \tau }{4}\right)}{8 \gamma ^3 \tau }
\end{equation}\\
\begin{equation}
    T_1\tilde{L}_{122}=T_2\tilde{L}_{211}=\frac{\left(4 \sinh \left(\frac{\gamma  \tau }{2}\right)-\gamma  \tau \right) \text{sech}^2\left(\frac{\gamma  \tau }{4}
\right)}{8 \gamma ^3 \tau }
\end{equation}

\begin{equation}
    T_1\tilde{L}_{112}=T_2\tilde{L}_{221}=\frac{\tanh \left(\frac{\gamma  \tau }{4}\right)}{\gamma ^3 \tau }
\end{equation}
for the linear contribution of $\alpha$. Similarly, for
linear drivings, their expressions are listed below:

\begin{equation}
T_1L_{11}=T_2L_{22}=\frac{\gamma  \tau  (\gamma  \tau-2 )+2 (4-\gamma  \tau ) \tanh \
\left(\frac{\gamma  \tau }{4}\right)}{8 \gamma ^3 \tau }\quad,\quad
    T_1L_{12}=T_2L_{21}=\frac{(\gamma  \tau-4 ) \tanh \left(\frac{\gamma  \tau }{4}\right)+\gamma  \tau }{4 \gamma ^3 \tau }    
\end{equation}
for Onsager ones and

\begin{equation}
T_1\tilde{L}_{111}=\frac{-\gamma ^2 \tau ^2+e^{\frac{3 \gamma  \tau }{2}} [\gamma  \tau  (\gamma  \tau -8)+24]+e^{\gamma  \tau } [\gamma  \tau  (\gamma  \tau -4)-24]-3 e^{\frac{\gamma  \tau }{2}} [\gamma  \tau  
(\gamma  \tau -4)+8]+24}{8 \gamma ^4 \tau  
\left(e^{\frac{\gamma  \tau }{2}}-1\right) \left(e^{\frac{\gamma  \tau }{2}}+1\right)^2},    
\end{equation}

\begin{equation}
     T_1\tilde{L}_{122}=T_2\tilde{L}_{211}=\frac{  \left(e^{\frac{\gamma  \tau }{2}}-1\right) \left[2 
e^{\frac{3 \gamma  \tau }{2}} (\gamma  \tau -4)+2 e^{\gamma  \tau } (\gamma  \tau +4)+e^{\frac{\gamma  \tau }{2}} [\gamma  \tau  (\gamma \tau -4)+8]-8\right]}{4 \gamma ^4 \tau  \left(e^{\gamma  \tau }-1\right)^2}
\end{equation}
and

\begin{equation}
 T_1\tilde{L}_{112}=T_2\tilde{L}_{221}=\frac{\left[\gamma  \tau +(\gamma  \tau -4) \tanh 
\left(\frac{\gamma  \tau }{4}\right)\right]}{4 \gamma ^4 \tau }   
\end{equation}
for the linear contribution in $\alpha$.

\end{document}